\documentclass[9pt,twocolumn,twoside]{opticajnl}
\journal{opticajournal} 

\setboolean{shortarticle}{true}

\hypersetup{
	colorlinks = true,
	urlcolor   = [RGB]{0,172,172},
	linkcolor  = [RGB]{0,172,172},
	citecolor  = [RGB]{0,172,172}
}
\renewcommand*{\phi}{\varphi}
\renewcommand*{\epsilon}{\varepsilon}
\DeclareMathOperator{\Tr}{Tr}
\DeclareMathOperator{\var}{var}
\usepackage{braket}
\usepackage{lineno}
\usepackage{bm}

\title{Error mitigated variational algorithm on a photonic processor}

\author[1,2,*]{O.V.~Borzenkova}
\author[3]{G.I.~Struchalin}
\author[3]{I.V.~Kondratyev}
\author[3]{A.D.~Moiseevskiy}
\author[3]{N.N.~Skryabin}
\author[1,3]{I.V.~Dyakonov}
\author[1,3]{S.S.~Straupe}

\affil[1]{Russian Quantum Center, 30 Bolshoy Boulevard, building 1, Moscow, 121205, Russia}
\affil[2]{Skolkovo Institute of Science and Technology, 3 Nobel Street, Moscow 121205, Russian Federation}
\affil[3]{Quantum Technology Centre and Faculty of Physics, M.\,V. Lomonosov Moscow State University, 1 Leninskie Gory, Moscow, 119991, Russia}

\affil[*]{oksana.borzenkova@skoltech.ru}

\begin{abstract}
Our study demonstrates successful error mitigation  of indistinguishability-related noise in a quantum photonic processor through the application of the zero-noise extrapolation technique. By measuring observable values at different error levels, we were able to extrapolate towards a noise-free regime. We examined the impact of partial distinguishability of photons in a two-qubit processor implementing the variational quantum eigensolver for a Schwinger Hamiltonian. Our findings highlight the effectiveness of the extrapolation technique in mitigating indistinguishability-related noise and improving the accuracy of Hamiltonian eigenvalue estimation. 
\end{abstract}

\setboolean{displaycopyright}{false}

\begin{document}

\maketitle

\textbf{Introduction.} Noise and errors are still the major obstacles for the development of scalable quantum computers. This problem, at least in theory, is remedied by quantum error correction (QEC). However, QEC requires additional overhead for the most valuable resources~--- the physical qubits and the circuit depth. The other way around could be careful account of experimentally observable noise processes. Recently multiple quantum error mitigation (QEM) methods were proposed: zero-noise extrapolation (ZNE) \cite{li2017efficient, krebsbach2022optimization}, quantum subspace expansion (QSE) \cite{mcclean2017hybrid, mcclean2020decoding}, purification \cite{o2022purification, yamamoto2022error},  probabilistic error cancellation (PEC) \cite{temme2017error}, symmetry verification \cite{mcardle2019error, PhysRevA.98.062339}, and continuous-time Markov-process error mitigation~\cite{barron2020measurement}. 

Experimental demonstrations of the QEM principle include implementation on superconducting processors \cite{temme2017error, kandala2019error,  BravyiPRA2021} and trapped ion systems \cite{zhang2020error, chen2023error}. Implementations of QEM on quantum photonic processors are of most relevance for this work. At first the PEC and ZNE methods were put under test to mitigate photon losses in a Gaussian boson sampler \cite{su2020error}. Later the PEC technique was used as a unified error mitigation scheme for any type of noise \cite{lee2022error}. It was experimentally applied to a variational quantum eigensolver (VQE) algorithm for a HeH$^+$ ion, where information was encoded in optical ququart states.

In this work, we apply the ZNE mitigation technique to the VQE algorithm which finds the minimal eigenvalue of the Schwinger Hamiltonian on a photonic quantum processor. A key source of errors in photonic architectures is the non-ideal indistinguishability of photons, which affects the Hong-Ou-Mandel (HOM) interference and detrimentally reduces the fidelity of linear-optical two-qubit gates. We demonstrate the power of the ZNE technique by extrapolating data gathered from a photonic processor with controllable indistinguishability towards the noise-free regime. 

\begin{figure}[h]
	\centering
	\includegraphics[width=0.95\linewidth]{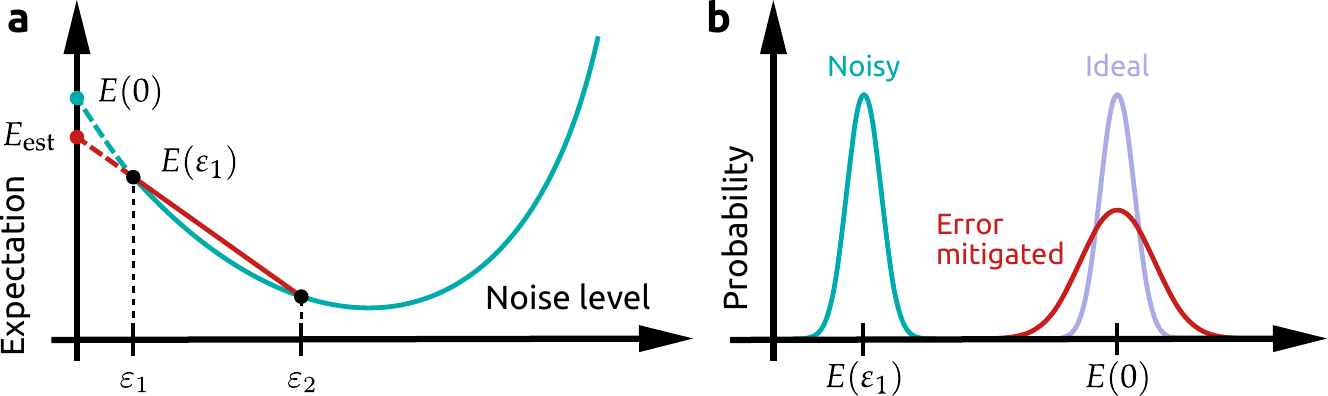}
	\caption{The idea of zero-noise extrapolation method. (a) Cyan curve describes the possible dependence of the expectation value $E(\epsilon)$ on the noise level $\epsilon$. Red line~--- linear extrapolation from two points with different noise levels. (b) Sketch of possible probability density function (PDF) of measured expectation value with noise (noisy, cyan), without noise (ideal, lilac), and ZNE estimator $E_\text{est}$ (error mitigated, red).}
	\label{fig:QEM_base}
\end{figure}

\begin{figure*}
\centering
\includegraphics[width=0.9\linewidth]{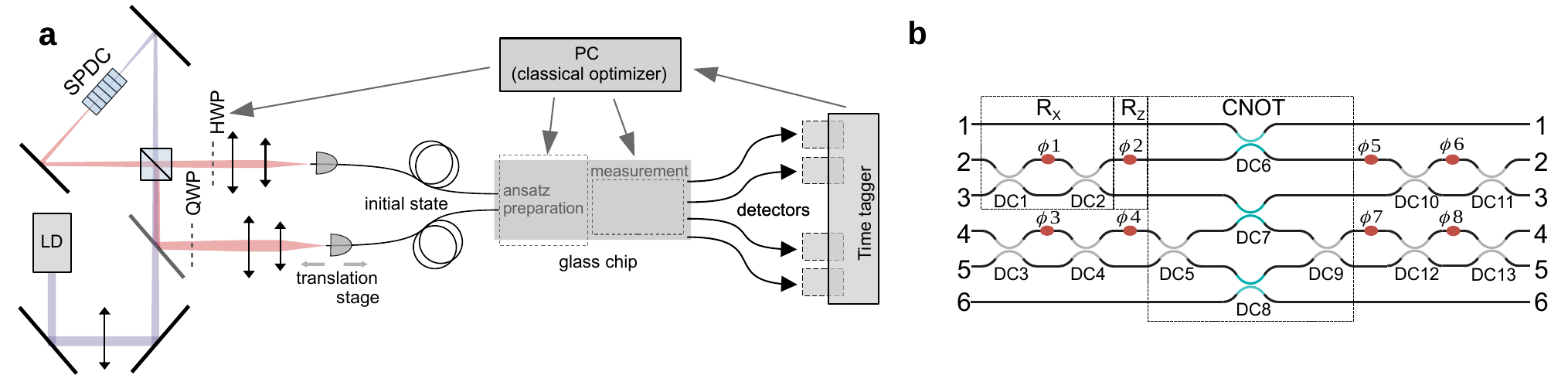}
\caption{(a) Experimental setup. Photon pairs generated in the nonlinear PPKTP crystal are coupled to the input of a programmable glass chip through optical fibers. The fibers at the output of the optical chip are coupled to single-photon superconducting detectors. The detection events are then processed by a time-tagger module and a classical computer. (b) Detailed scheme of the integrated interferometer used in the quantum processor. The interferometer is formed by directional couplers (grey and cyan ones have the splitting ratio of 50:50 and 33:67, respectively) and programmable thermo-optic phase-shifters (red ovals).} 
\label{fig:experimental setup}
\end{figure*}

\textbf{Zero-noise extrapolation.} Suppose a quantum system is characterized by a \emph{controllable} parameter $\epsilon \ge 0$ that determines the noise or decoherence strength (e.\,g., photon distinguishability). We call $\epsilon$ the \emph{noise level}. Let $\epsilon = 0$ correspond to an ideal case where all undesirable noise is absent. A common task for ZNE is noise suppression when measuring, e.\,g., expectation value $E = \braket{H} = \Tr H \rho$ of some observable $H$. The state $\rho(\epsilon)$ produced by an experimental setup depends on $\epsilon$ hence the expectation $E(\epsilon)$. The aim of ZNE is to find an estimation $E_\text{est} \approx E(0)$ given only a measured data set for $0 < \epsilon_1 < \dots < \epsilon_n$:
\begin{equation}
    E_\text{est} = F \bigl( E(\epsilon_1), \dots, E(\epsilon_n); \epsilon_1, \dots, \epsilon_n \bigr), \label{eq:ZNECommon}
\end{equation}
The lowest possible value $\epsilon_1$ corresponds to intrinsic unavoidable noise in the experimental setup. The approximating function $F$ gives the rule of extrapolation.

In our experiments the noise level $\epsilon$ is related to HOM interference visibility in the optical chip. In this case, as shown in Supplemental Material, the dependence $E(\epsilon)$ for $\epsilon \ll 1$ is well approximated by a linear function:
\begin{equation} 
	E(\epsilon) = c_1 + c_2 \epsilon,
\end{equation}
where $c_1$ and $c_2$ are unknown coefficients. For linear dependence, measurements at two points $\epsilon_{1,2}$ are sufficient to estimate $c_{1,2}$. Therefore, the general ZNE form~\eqref{eq:ZNECommon} reduces to
\begin{equation}
	E_\text{est} = \frac{\epsilon_2 E(\epsilon_1)  - \epsilon_1 E(\epsilon_2)}{\epsilon_2 - \epsilon_1}.  \label{eq:LinearEstimator}
\end{equation}
While the estimation $E_\text{est}$ is closer to $E(0)$ than $E(\epsilon_1)$ (for a proper selected extrapolation function), this improvement usually comes at a price of the increased estimator variance $\var[E_\text{est}]$ (see Fig.~\ref{fig:QEM_base}). For example, if the statistical uncertainty is the same for each $E(\epsilon_{1,2})$ regardless of noise, $\var[E(\epsilon_1)] = \var[E(\epsilon_2)] = \sigma^2$, the variance is
\begin{equation}
    \var[E_\text{est}] = \sigma^2 \frac{\epsilon_2^2 + \epsilon_1^2}{(\epsilon_2 - \epsilon_1)^2}. \label{eq:EstimatorVariance}
\end{equation}
Note that $\var[E_\text{est}]$ may be arbitrarily large as $\epsilon_2$ approaches $\epsilon_1$: $\epsilon_2 \to \epsilon_1$. However, a special choice of measured levels $\{\epsilon_1, \dots, \epsilon_n\}$ allows to build advanced extrapolation with reduced estimator variance~\cite{krebsbach2022optimization}.

\textbf{Variational eigensolver.} We verify the efficiency of ZNE in the VQE experiment. The VQE algorithm aims to find the ground energy $E_0$ of a given Hamiltonian $H$~\cite{tilly2022variational}. A quantum processor prepares a probe state $\ket{\psi(\bm \phi)}$ by means of an \emph{ansatz} circuit that depends on some adjustable parameters $\bm \phi$. This state is measured, and the expectation value $\braket{H} = \braket{\psi(\bm \phi) | H | \psi(\bm \phi)}$ is calculated using the measurement results. This value $\braket{H}$ is limited by the the ground energy $E_0$:
\begin{equation}
    \braket{\psi(\bm \phi) | H | \psi(\bm \phi)} \ge E_0. \label{eq:VQE}
\end{equation}
For sufficiently expressible ansatz, the minimum of $\braket{H}$ over $\bm \phi$ is equal to $E_0$. Thus, in what follows, we will use terms ``ground energy'' and ``eigenvalue'' interchangeably.

The search over the parameter space $\{\bm \phi\}$ is the task of a classical optimization algorithm. We have selected the simultaneous perturbation stochastic approximation (SPSA) method~\cite{Bhatnagar2013}, since it is tolerant to random perturbations in target-function values. During the VQE optimization run, the mitigated quantity is the measured expectation value $\braket{H}$ that is forwarded to the optimizer. As the result of the VQE algorithm we obtain the estimation $E_\text{est}$ of the Hamiltonian minimal eigenvalue.

In the present work we study error mitigation and not a specific Hamiltonian, so we choose the Schwinger model Hamiltonian~\cite{PhysRevD.66.013002} which we previously analyzed in the similar VQE setup \cite{BorzenkovaAPL21}. We believe that this Hamiltonian is quite general, and conclusions about ZNE remain valid for other choices. For two qubits the Schwinger Hamiltonian has the following form:
\begin{equation}
	H(m) = I + \sigma_1^x \sigma_2^x + \sigma_1^y \sigma_2^y - \frac{1}{2} \sigma_1^z + \frac{1}{2} \sigma_1^z \sigma_2^z + \frac{m}{2} (\sigma_2^z - \sigma_1^z),
\end{equation}
where $I$ is the identity matrix, $\sigma_i$ ($i = 1, 2$) are the Pauli matrices acting on the qubit $i$, and $m \in \mathbb R$ is the Hamiltonian parameter.

\textbf{Experimental setup.}
We utilize a source of photon pairs based on spontaneous parametric down-conversion (SPDC) and introduce controllable distinguishability by changing the polarization of one photon in a pair relative to the other.

Our two-qubit photonic processor consists of integrated optical interferometer manufactured in fused silica chip by femtosecond laser writing \cite{skryabin2023two}. The photons from SPDC source are injected to the chip through polarization-maintaining single-mode optical fibers and outcoupled using regular single-mode fibers connected to superconducting single-photon detectors (see Fig.~\ref{fig:experimental setup}a). Photons at the 810\,nm wavelength are generated by a nonlinear 30-mm PPKTP crystal that is pumped by a 405-nm CW laser diode in the Sagnac configuration. The photon pairs generation rate is 150\,kHz. The SPDC source may generate polarization-entangled photons when pumping the crystal from both sides, but it is not necessary in the present experiment. So the source was adjusted to produce the factorized state $\ket{HH}$, when both photons from a pair have horizontal polarization.

The elementary blocks of the programmable photonic chip are  phase-shifters (PS) and directional couplers (DC), which implement reconfigurable single-qubit gates and a two-qubit CNOT gate (see Fig.~\ref{fig:experimental setup}b). The first part of the interferometer (phase-shifters 1--4) prepares the probe state, and the right part (phase-shifters 5--8) sets the measurement basis. The central area of the chip consists of passive DCs, which implement the probabilistic linear optical CNOT gate with $1/9$ success probability \cite{ralph2002linear}.

By definition we set the noise level $\epsilon := 1 - V$, where $V$ is the HOM visibility inside the chip. We can measure~$V$ by setting the phase $\phi_4$ to $\pi$ and examining the coincidence counts between outputs 3 and 5. The interference happens at the central directional coupler DC7. Since DC7 is not ideal, the experimentally measured HOM dip visibility in the chip drops to $82 \%$ compared with $98 \%$ in the fiber splitter.

\begin{figure}
\centering
{
    \includegraphics[width=0.7\linewidth]{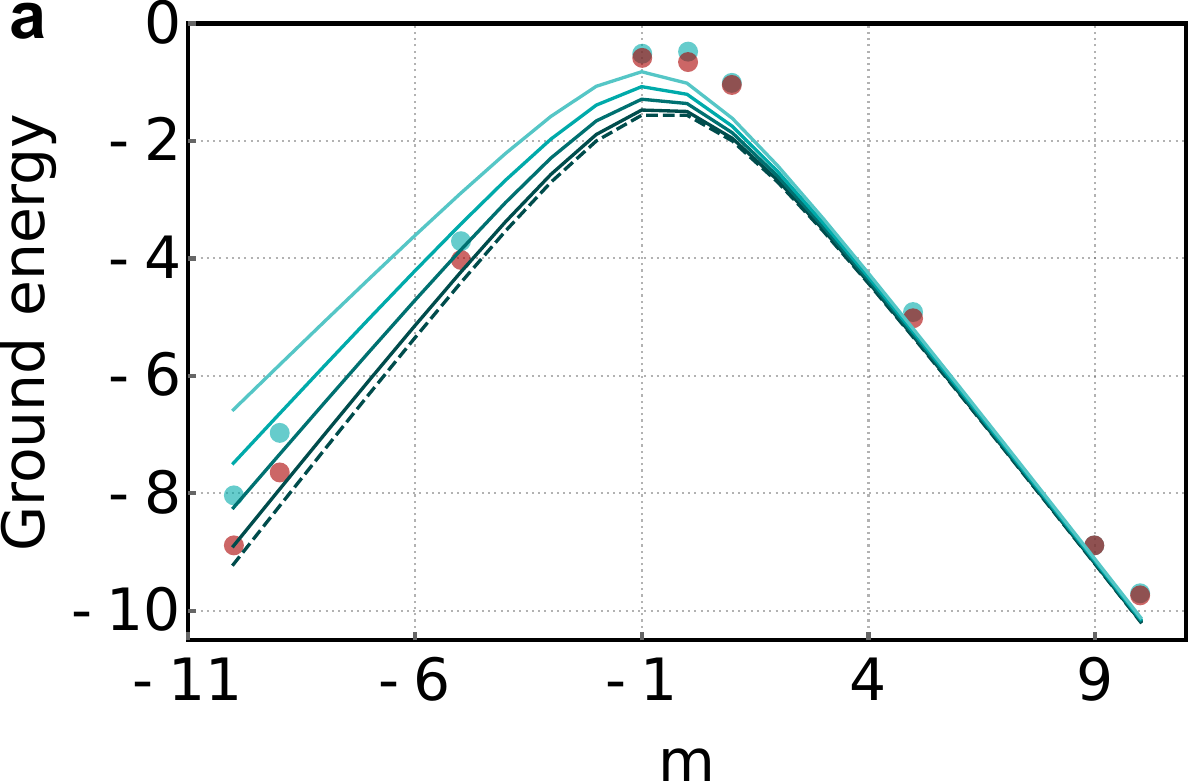}
    \label{fig:evall}
}\vspace{0.25cm}
{
    \includegraphics[width=0.7\linewidth]{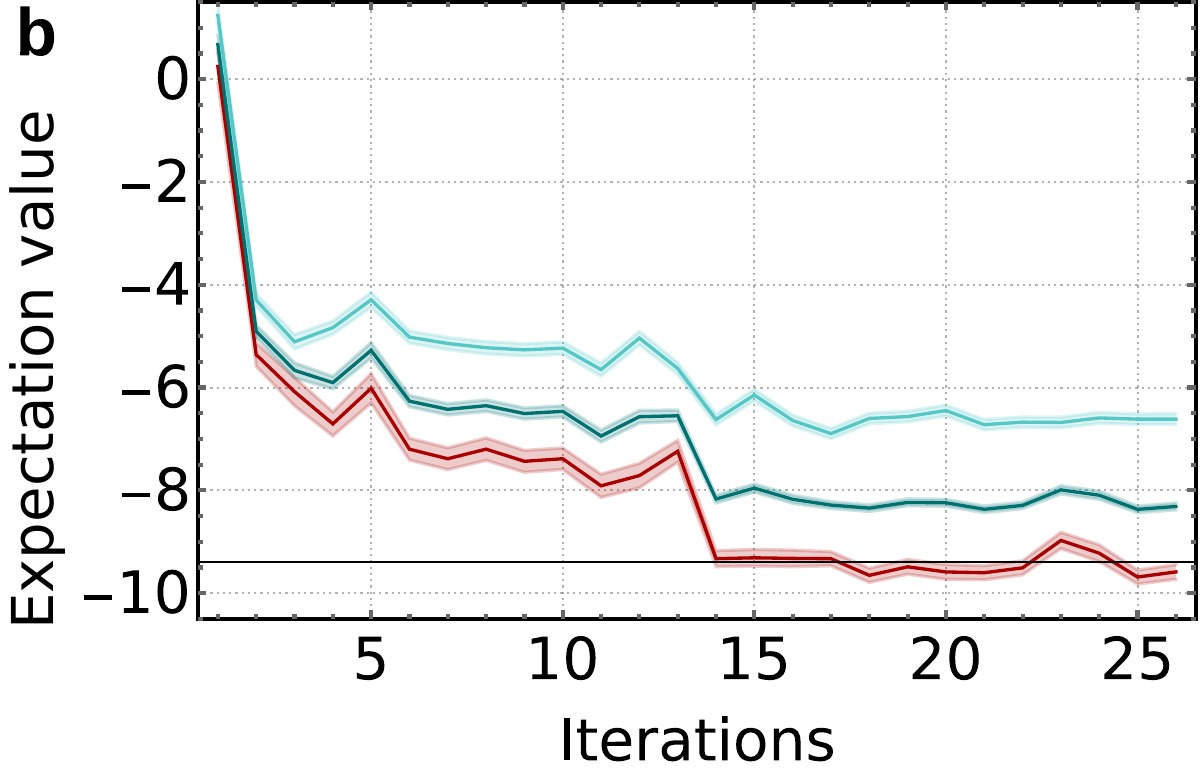}
    \label{fig:convergence_m10}
}
\caption{(a) Dependence of ground energy $E$ found by VQE for the Schwinger Hamiltonian $H(m)$ on different $m$ and variable noise level $\epsilon_1$. Dashed line~--- exact theoretical ground energy $E_0$; four cyan lines (bottom-up)~--- numerical simulations for $\epsilon_1 = 0.1, 0.3, 0.5, 0.7$; cyan and red dots~--- experiment without and with mitigation, respectively. (b) Experimental dependence of the Hamiltonian expectation value $\braket{H(-10)}$ on the number of SPSA-optimizer iterations during a single VQE run. Cyan curves~--- measured expectations $E(\epsilon_1)$ and $E(\epsilon_2)$ for noise levels $\epsilon_{1,2}$; red-curve~--- corresponding error-mitigated result $E_\text{est}$; black horizontal line~--- exact ground energy $E_0$; filling~--- one standard deviation of the appropriate quantity under assumption that photon-counting statistics is Poissonian.} 
\label{fig:exp}
\end{figure}

We alter the noise level $\epsilon$ by rotating the polarization of one photon from a pair thus changing indistinguishability of photons and HOM visibility. A half-wave plate (HWP) installed in one arm of the SPDC source transforms the polarization state $\ket{H}$ to $\cos(2\theta) \ket{H} + \sin(2\theta) \ket{V}$, where $\theta$ is the HWP axis angle relative to the horizontal.

The intrinsic noise $\epsilon_1 = 0.18$ corresponds to the case when the experimental setup is adjusted to give the highest possible visibility $V_1 = 82\%$. The second value $\epsilon_2 = 0.29$ is obtained from visibility measurement, $V_2 = 71\%$, when HWP is rotated by $10^\circ$ from its optimal position.

\textbf{ZNE on a photonic processor.} We run VQE algorithm for each value of the Hamiltonian parameter $m$, first, without mitigation with intrinsic noise level $\epsilon_1$ and, second, with error mitigation at noise levels $\epsilon_{1,2}$.

Numerical simulations and experimental results show that the expectation $\braket{H(m)}$ is mainly sensitive to noise in the region of negative $m$, as shown in Fig.~\ref{fig:exp}. 

ZNE mitigation can significantly improve the quality of ground-energy estimation by VQE. It is possible to obtain values closer to the precise eigenvalues $E_0$ and below the limit set by the intrinsic distinguishability of the photon source. 
For example, without mitigation with intrinsic noise, the VQE algorithm converges to $E = -7.9\pm 0.1$ for $m = -10$, while with mitigation we obtain $E = -8.9\pm 0.6$ (the precise value is $E_0 = -9.2$).

\begin{figure}
	\centering
	\includegraphics[width=0.8\linewidth]{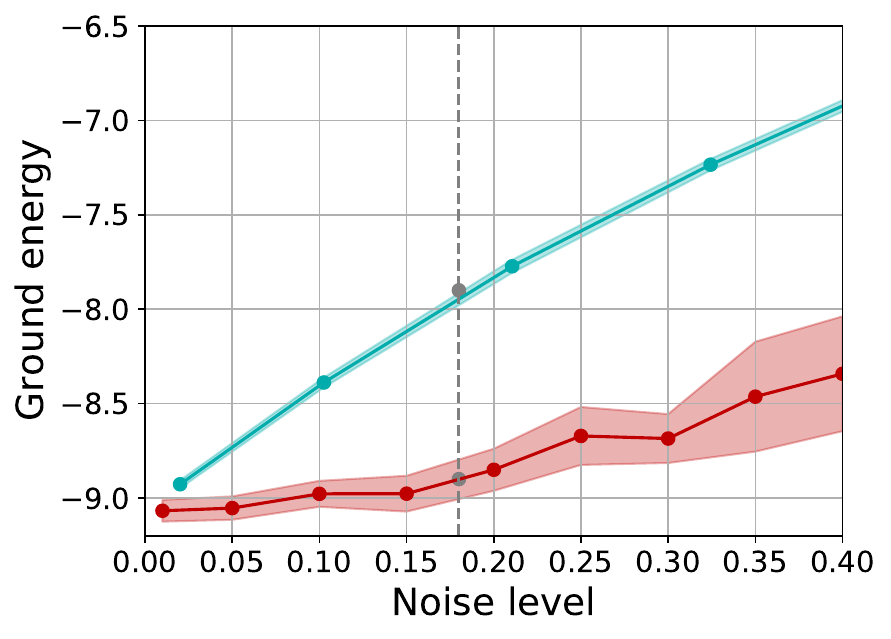}
	\caption{Dependence of ground energy $E$ found by VQE for the Schwinger Hamiltonian $H(-10)$ on the intrinsic noise level $\epsilon_1$. Cyan curve~--- simulation without mitigation; red curve~--- simulation with mitigation, where the second noise level is $\epsilon_2 = t \epsilon_1$ and the ratio $t \approx 1.61$ as in the experiment; filling~--- one standard deviation of mean over 100 optimizer runs starting from random initial guesses; gray points~--- experimental results.}
	\label{fig:EigenVsIndis}
\end{figure}

In numerical simulations, we examined how error mitigation works for different noise levels $\epsilon$ in more detail to assess the method's applicability to larger distinguishability. Each point in Fig.~\ref{fig:EigenVsIndis} is the ground-energy estimation found by the SPSA optimizer after 200 iterations averaged over 100 runs starting from random phases $\{\phi_1, \dots, \phi_4\}$. For a relatively poor indistinguishability $V = 60\%$ ($\epsilon = 0.4$), ZNE allows us to achieve better results than our processor without error mitigation.

\textbf{Deferred mitigation.}
A typical problem in optical experiments is the unstable system efficiency during long-term measurements due to temperature and mechanical drifts. To speedup VQE we propose to turn on the mitigation not from the very beginning but only after a certain iteration of the optimizer, so the first $k_0$ iterations are not mitigated and subsequent $k_1$ ones are mitigated. We refer to this strategy as a \emph{deferred mitigation}. We show below that for fixed VQE-experiment time deferred mitigation ($k_0 > 0$) achieves higher accuracy than ordinary strategy ($k_0 = 0$). Equivalently, for given accuracy, deferred mitigation is faster.

Experiment time is proportional to the total number of measured bases~$N$ needed for estimation of expectation values $\braket{H}$ during VQE optimization. Suppose that without mitigation each optimizer iteration requires $n$ basis measurements ($n = 6$ for the Schwinger Hamiltonian and the SPSA optimizer, see Supplemental Material). When the mitigation is applied this quantity doubles to $2n$ bases per iteration because two noise levels~$\epsilon_{1,2}$ must be accounted. Therefore, at iteration $K = k_0 + k_1$ of the optimizer, the number of performed measurements is
\begin{equation}
	N = n k_0 + 2 n k_1 = n (K + k_1), \label{eq:NumberOfMeasurements}
\end{equation}
where $0 \le k_{0,1} \le K$ is by definition. If $N$ is fixed, there is a trade-off between optimizer convergence (parameter $K$) and noise mitigation (parameter $k_1$).

We quantify VQE error $\Delta E = |E - E_0|$ by the absolute difference between ground-energy estimation $E$ found by VQE and the exact theoretical value $E_0$. Two factors affect this error: first, deviation of ansatz parameters $\{\phi_j\}$ from their optimal values and, second, nonzero noise $\epsilon$. At the beginning of VQE optimization, usually the first factor prevails, and the mitigation has little impact on accuracy since it aims to suppress the second one. Therefore, ZNE may be turned off not to waste measurement budget~$N$. As the optimizer refines parameters $\{\phi_j\}$, the noise factor becomes stronger, and the mitigation comes into play.

\begin{figure}
	\centering
	\includegraphics[width=0.7\linewidth]{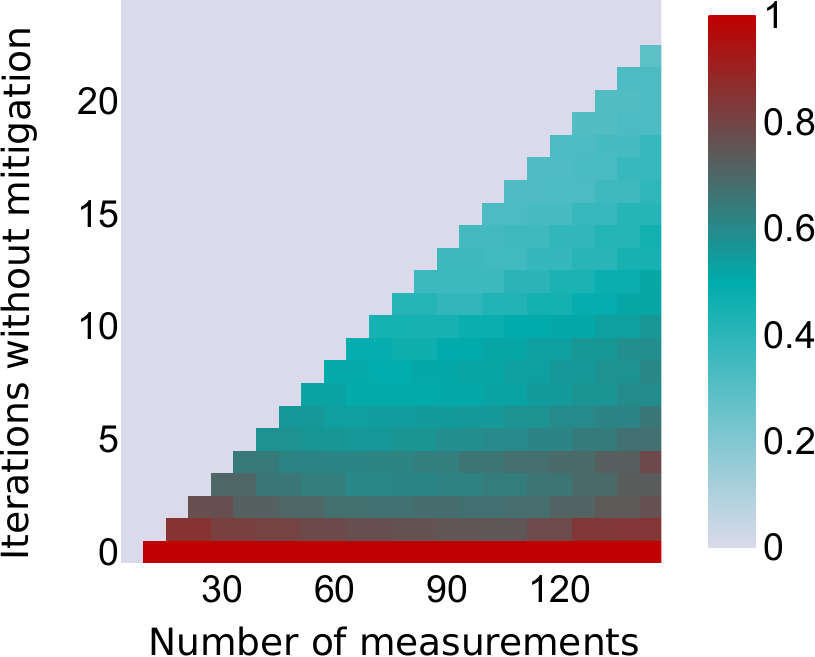}
	\caption{Simulation of deferred mitigation. Mitigation is turned on after an iteration $k_0$ indicated by the y-axis, while the number of measured bases $N$ is shown by the x-axis. The color scale indicates relative error $R(N, k_0)$ of VQE (see text for definition).}
	\label{fig:wo+w_mitigation}
\end{figure}

We conducted numerical simulations of the experiment for different values $k_0$. For each $k_0$ we performed 100 VQE runs, tracked the number $N$, and calculated error $\Delta E(N, k_0) = |\bar E(N, k_0) - E_0|$, where $\bar E(N, k_0)$ is the averaged-over-runs ground energy found by the optimizer utilizing $N$ measurements. In our implementation of deferred mitigation, SPSA optimizer is restarted from scratch when mitigation is switched on, using the latest probe state found in unmitigated stage as the initial guess.

The obtained dependence of \emph{relative error} $R(N, k_0) := \Delta E(N, k_0) / \Delta E(N, 0)$ is depicted in Fig.~\ref{fig:wo+w_mitigation}. By definition $R(N, 0) = 1$, and values $R(N, k_0) < 1$ mean that deferred mitigation achieves better accuracy (error is smaller) compared to ordinary mitigation with $k_0 = 0$, given the same number of measurements $N$. As one can see, for fixed $N$ and $k_0 > 0$, the error $R(N, k_0)$ is always below unity and decreases towards larger $k_0$, e.\,g., $R(120, 18) \approx 0.194$.
Therefore, for this setting deferred mitigation is always preferable and the highest accuracy is achieved when only the last optimizer iteration is mitigated: $k_1 = 1, k_0 = N/n - 2$.

\textbf{Conclusion.}
We have experimentally demonstrated ZNE method to mitigate partial distinguishability of photons in linear-optical quantum processors. We have studied the performance of the method  
in a two-qubit VQE algorithm and have found that error mitigation allows to estimate the ground energy to better accuracy than allowed by intrinsic photon source distinguishability. 
For higher dimensional systems the dependence of a mitigated quantity on the noise level may become nonlinear. Nevertheless, functional form of this dependence (e.\,g. degree of an approximating polynomial) can still be calculated analytically or measured experimentally. ZNE, along with other QEM methods, has several limitations \cite{takagi2022fundamental}, including a longer experiment time and an increase in estimator variance; both of these parameters can be assessed beforehand to choose a more effective QEM strategy. We will address ZNE implementation in larger scale  experiments in future works.

Partial photon distinguishability is ubiquitous for photon sources of different physical nature, so we expect that ZNE mitigation may be widely adopted in linear-optical quantum computing to overcome corresponding errors regardless of the specific realization of two-qubit gates and quantum algorithms. The method only requires a possibility to change the degree of distinguishability in the experiment. The specific approach using polarization control demonstrated here will be applicable for any architecture with dual-rail encoding of photonic qubits in waveguides or spatial modes.

\begin{backmatter}

\bmsection{Acknowledgments} The authors acknowledge support by Rosatom in the framework of the Roadmap for Quantum computing (Contract No. 868-1.3-15/15-2021 dated October 5, 2021 and Contract No.P2154 dated November 24, 2021). I.D. acknowledges support from  Russian Science Foundation grant 22-12-00353 (https://rscf.ru/en/project/22-12-00353/).

\bmsection{Disclosures} The authors declare no conflicts of interest.

\bmsection{Data availability} Data underlying the results presented in this paper are not publicly available at this time but may be obtained from the authors upon reasonable request.

\bmsection{Supplemental document} See Supplemental Material for supporting content.

\end{backmatter}

\bibliography{biblio}
\bibliographyfullrefs{biblio}

\end{document}


\maketitle

\section{Measurement of Hamiltonian expectation value}
Let us remind two-qubit Schwinger Hamiltonian $H$ used in the main article:
\begin{equation}
	H = I + \sigma_1^x \sigma_2^x + \sigma_1^y \sigma_2^y - \frac{1}{2} \sigma_1^z + \frac{1}{2} \sigma_1^z \sigma_2^z + \frac{m}{2} (\sigma_2^z - \sigma_1^z).
\end{equation}
The experimental setup cannot directly measure the Hamiltonian expectation value $\braket{H} = \braket{\psi | H | \psi}$, since it performs only projective measurements onto vectors of some orthonormal basis. The Hamiltonian $H$ is expressed as a sum of Pauli strings $S_i$ with some real coefficients $h_i$: $H = \sum_i h_i S_i$. Its expectation value $\braket{H}$ also has this form: $\braket{H} = \sum_i h_i \braket{S_i}$. Thus, having measured individual terms $\braket{S_i}$, the expectation $\braket{H}$ is calculated by summation. In turn, any Pauli-string expectation $\braket{S_i}$ can be estimated via outcomes of a projective measurement in eigenbasis of $S_i$. For example, to measure $\braket{\sigma_1^z \sigma_2^z}$ one carries out measurements in its eigenbasis $\{\ket{00}$, $\ket{01}$, $\ket{10}$, $\ket{11}\}$, obtaining accumulated counts $n_{00}^z, \dots, n_{11}^z$ that are proportional to corresponding Born-rule probabilities $p_{00}^z, \dots, p_{11}^z$:
\begin{equation}
	\langle \sigma_1^z \sigma_2^z \rangle = p_{00}^z - p_{01}^z - p_{10}^z + p_{11}^z \approx \frac{n_{00}^z - n_{01}^z - n_{10}^z + n_{11}^z}{n_{00}^z + n_{01}^z + n_{10}^z + n_{11}^z}. \label{eq:ZZEstimation}
\end{equation}

Observe that all terms in $H$ can be divided into 3 groups $\mathfrak S_j$ of mutually commuting Pauli strings:
\begin{equation}
	\mathfrak S_1 = \{\sigma_1^x \sigma_2^x\}, \quad
	\mathfrak S_2 = \{\sigma_1^y \sigma_2^y\}, \quad
	\mathfrak S_3 = \{I,\ \sigma_1^z,\ \sigma_2^z,\ \sigma_1^z \sigma_2^z\}.
\end{equation}
Pauli-string expectations in each group $\mathfrak S_j$, $j = 1, 2, 3$, can be estimated by measuring the probe state $\ket{\psi}$ in eigenbases of $\sigma_1^x \sigma_2^x$, $\sigma_1^y \sigma_2^y$, and $\sigma_1^z \sigma_2^z$, respectively. So 3 basis measurements are required for estimation of $\braket{H}$.

Altogether, we arrive to the following formula for the Hamiltonian expectation value $\braket{H}$:
\begin{align}
	\braket{H} =\
	& p_{00}^x - p_{01}^x - p_{10}^x + p_{11}^x + \nonumber \\
	& p_{00}^y - p_{01}^y - p_{10}^y + p_{11}^y + \nonumber \\
	& p_{00}^z - m p_{01}^z + (1+m) p_{10}^z + 2 p_{11}^z, \label{eq:HamiltonianFromProb}
\end{align}
where $p_{ij}^k$ ($i, j = 0, 1$, $k = x, y, z$) is the Born-rule probability of an outcome $ij$ when measuring $\ket{\psi}$ in the eigenbasis of $\sigma_1^k \sigma_2^k$. Each $p_{ij}^k$ is estimated using accumulated counts $n_{ij}^k$ in complete analogy to~\eqref{eq:ZZEstimation}. It is important to note that $\braket{H}$ is a linear function of outcome probabilities~$p_{ij}^k$.

SPSA optimizer evaluates the target function $\braket{H}$ two times per iteration, consequently, it spends 6 basis measurements per iteration.

\section{Dependence of noise level on half-wave plate axis angle}
By definition we connect the noise level $\epsilon$ with Hong-Ou-Mandel (HOM) visibility $V$: $\epsilon = 1 - V$. In the experimental setup visibility $V$ is altered by rotating polarization of one photon from a pair via half-wave plate (HWP). In this section we calculate the dependence of $\epsilon(\theta)$ on HWP-axis angle $\theta$.

Suppose the photon source produces pairs where each photon has horizontal polarization, then HWP transforms the state of the pair:
\begin{equation}
	a_1^H a_2^H \vac \to (a_1^H \cos{(2\theta)} + a_1^V \sin{(2\theta)}) a_2^H \vac, \label{eq:HWPTransform}
\end{equation}
where $a_i^P$, $i = 1,2$, $P = H, V$ is the photon-birth operator in mode $i$ with polarization $P$, and $\vac$ is the vacuum state. To measure HOM visibility, the photons are send to input ports of a symmetrical directional coupler (DC), so input birth operators $a$ are expressed via output ones $b$ as
\begin{equation}
	a_1^P = \frac{b_1^P + b_2^P}{\sqrt{2}}, \quad a_2^P = \frac{b_1^P - b_2^P}{\sqrt{2}}. \label{eq:DCBirthOps}
\end{equation}
By substituting~\eqref{eq:DCBirthOps} into~\eqref{eq:HWPTransform}, we find the state $\ket{\psi_\text{out}(\theta)}$ after DC:
\begin{equation}
	\ket{\psi_\text{out}(\theta)} = \left[\frac{(b_1^H)^2 - (b_2^H)^2}{2} \cos{(2\theta)} + \frac{b_1^V b_1^H - b_2^V b_2^H + b_2^V b_1^H - b_1^V b_2^H}{2} \sin{(2\theta)} \right] \vac. \label{eq:StateAfterDC}
\end{equation}

We measure coincidence count rate between output modes 1 and 2 with detectors that are not sensitive to polarization, so the probability of coincidence $p(\theta)$ is:
\begin{equation}
	p(\theta) = \sum_{P_1, P_2 = H, V} | \braket{\text{vac} | (b_1^{P_1})^\dagger (b_2^{P_2})^\dagger | \psi_\text{out}(\theta)} |^2. \label{eq:ProbOnThetaGeneral}
\end{equation}
Considering canonical commutation relations between birth operators $b$, $p(\theta)$ simplifies to:
\begin{equation}
	p(\theta) = \frac{\sin^2{(2\theta)}}{2}.
\end{equation}
When $\theta = 0$ the photons are fully indistinguishable, and HOM-dip magnitude is maximal~--- the probability is zero, $p(0) = 0$, and HOM visibility $V = 1$. For completely distinguishable photons, coincidence probability is $1/2$, and $V = 0$. Therefore, HOM visibility $V$ is defined as:
\begin{equation}
	V(\theta) = \frac{1/2 - p(\theta)}{1/2 + p(\theta)}.
\end{equation}
By setting the noise level $\epsilon = 1 - V$, we arrive to the explicit dependence $\epsilon(\theta)$ after elementary transformations:
\begin{equation}
	\epsilon(\theta) = 2 \frac{\sin^2(2\theta)}{1 + \sin^2(2\theta)}. \label{eq:EpsilonOnTheta}
\end{equation}

\section{Dependence of Hamiltonian expectation value on noise level}

In this section we prove that the dependence of the Hamiltonian expectation value $\braket{H}$ on the noise level $\epsilon$ is well approximated by a linear function, $\braket{H} \approx c_1 + c_2 \epsilon$, for $\epsilon \ll 1$ in our experiment. First, we calculate the dependence of any outcome probability $p(\theta)$ on the HWP angle $\theta$. Then, inverting the dependence $\epsilon(\theta)$~\eqref{eq:EpsilonOnTheta}, we show that $p(\epsilon) \equiv p(\theta(\epsilon))$ is approximately linear in $\epsilon$. Finally, since the expectation $\braket{H}$~\eqref{eq:HamiltonianFromProb} is a linear combination of outcome probabilities $p(\epsilon)$, the expectation $\braket{H}$ is also linear in $\epsilon$.

Let us consider a linear-optical chip with multiple input and output modes (in our experiment we use the chip with 6 modes). Without loss of generality, we calculate the probability of detecting a photon pair at output modes 1 and 2, when initially the pair is coupled to inputs 1 and 2. If input and output modes are not equal to 1 and 2, we simply renumber the modes leaving the expressions below the same. Suppose, the chip is manufactured in such a way that it does not affect polarization of photons. Again, assume the input state is described by~\eqref{eq:HWPTransform}. The chip performs some unitary transformation $U$ of modes, therefore, analogous to~\eqref{eq:DCBirthOps}, the input birth operators $a_i^P$ can be represented by output one $b_i^P$:
\begin{equation}
	a_i^P = \sum_j U_{ij} b_j^P, \label{eq:ChipBirthOps}
\end{equation}
where $U_{ij}$ are matrix elements of $U$. We substitute the transformation~\eqref{eq:ChipBirthOps} into the input state~\eqref{eq:HWPTransform} to find the output $\ket{\psi_\text{out}(\theta)}$:
\begin{equation}
	\ket{\psi_\text{out}(\theta)} = \sum_{ij} \left[
	U_{1i} U_{2j} b_i^H b_j^H \cos(2\theta) +
	U_{1i} U_{2j} b_i^V b_j^H \cos(2\theta)
	\right] \vac.
\end{equation}
The general expression~\eqref{eq:ProbOnThetaGeneral} for outcome probability $p(\theta)$ remains valid, and we get
\begin{equation}
	p(\theta) = |U_{11} U_{22} + U_{12} U_{21}|^2 \cos^2(2\theta) + (|U_{11} U_{22}|^2 + |U_{12} U_{21}|^2) \sin^2(2\theta). \label{eq:ProbOnThetaChip}
\end{equation}

By inverting the dependence $\epsilon(\theta)$~\eqref{eq:EpsilonOnTheta}, we obtain $\sin^2(2\theta) = \epsilon/(2-\epsilon)$, and after elementary transformations we find the dependence of the outcome probability $p(\epsilon)$ on the noise level~$\epsilon$:
\begin{equation}
	p(\epsilon) = |U_{11} U_{22} + U_{12} U_{21}|^2 - \frac{2 \epsilon}{2 - \epsilon} \Re (U_{11} U_{22} U_{12}^* U_{21}^*). \label{eq:ProbOnEpsilonChip}
\end{equation}
Any outcome probability contained in the Hamiltonian expectation value $\braket{H}$~\eqref{eq:HamiltonianFromProb} has similar functional form on $\epsilon$ as~\eqref{eq:ProbOnEpsilonChip}, only matrix elements of $U$ would change. Consequently, absorbing terms that does not depend on $\epsilon$ into constants $c_1$ and $c_2$, we arrive at the following expression:
\begin{equation}
	\braket{H} = c_1 + \frac{2\epsilon}{2 - \epsilon} c_2.
\end{equation}
Indeed, for small noises $\epsilon \ll 1$, denominator variation is negligible, and $\braket{H}$ is approximately linear in $\epsilon$: $\braket{H} \approx c_1 + \epsilon c_2$.

If the number of photons is more than two, a useful general approach for calculating visibility, expectation values, \textit{etc.} for partially distinguishable bosons or fermions can be found in Ref.~\cite{dittel2021wave}.

\bibliography{biblio}